\documentstyle[aps,preprint]{revtex}
\begin{document}
\draft{}
\title{ THE $\rho^{\pm}-\rho^0$ MASS SPLITTING PROBLEM}
\author{M.N. Achasov\\
G.I. Budker Institute for Nuclear Physics,\\
630090 Novosibirsk 90, Russia
\thanks{Electronic address: achasov@inp.nsk.su}
\\
and\\
N.N. Achasov \\
S.L. Sobolev Institute for Mathematics,\\
630090 Novosibirsk 90, Russia
\thanks{Electronic address: achasov@math.nsc.ru}}
\date{\today}
\maketitle
\begin{abstract}
It is discussed  the problem of the $\rho^\pm-\rho^0$ mass splitting.
It is suggested to use the $\phi\to\rho\pi\to 3\pi$ decay to measure the
$\rho^{\pm}-\rho^0$ mass splitting.
\end{abstract}


In the framework of the $SU(3)$ theory with the $U$-spin invariance of
electromagnetic interactions, taking into account the ideal $\omega-\phi$
mixing and ignoring $\rho^0-\phi$ mixing for the Okubo-Zweig-Iizuki (OZI)
rule reasons, it was obtained \cite{gourdin-67} for the $\rho^0-\omega$
mixing
\begin{eqnarray}
\label{uspinV}
&& - Re\left (\Pi_{\rho^0\omega}\right )= \left ( m^2_{K^{\ast\pm}}-
m^2_{K^{\ast 0}}\right )-\left ( m^2_{\rho^{\ast\pm}}-
m^2_{\rho^{\ast 0}}\right )\,.
\end{eqnarray}

The advent of quantum chromodynamics (QCD) did not affect Eq. (\ref{uspinV})
for the $U$-spin invariance of isospin symmetry breaking interactions
was not affected. But, now we perceived the importance of the $u-d$ quark
mass splitting in the isospin symmetry breaking, see, for example, review
\cite{gasser-82}. Eq. (\ref{uspinV}) is correct to terms caused by both
isospin symmetry breaking interactions and $SU(3)$ symmetry breaking
interactions ("semi-strong interactions"). It means that corrections up to
25\% to Eq. (\ref{uspinV}) are possible  .

Particle Data Group \cite{pdg-98} gives for the $K^{\ast\pm}-K^{\ast 0}$ mass
splitting
\begin{eqnarray}
\label{kstar0kstarpm}
&& m_{K^{\ast 0}} - m_{K^{\ast\pm}} = 6.7 \pm 1.2 \ \ \mbox{MeV}\,,
\end{eqnarray}
and for the $\rho^\pm - \rho^0$ mass splitting
\begin{eqnarray}
\label{rho0rhopm}
&& m_{\rho^0} - m_{\rho^\pm} = 0.1 \pm 0.9 \ \ \mbox{MeV}\,.
\end{eqnarray}
But the $\rho^\pm - \rho^0$ mass splitting can be calculated with Eq.
(\ref{uspinV}) taking into account the well specified $\omega\to\pi^+\pi^-$
decay \cite{pdg-98}.

Really, as was first pointed by Glashow \cite{glashow-61} the $\omega$ meson
decays into $\pi^+\pi^-$ via the $\rho^0-\omega$ mixing, see also , for
example, \cite{goldhaber-69,gourdin-69,renard-70,achasov-78,achasov-92},
\begin{eqnarray}
\label{omegato2pi}
&&B(\omega\to\pi^+\pi^-)=\frac{\Gamma \left (\rho^0\to\pi^+\pi^-\,;\,m_\omega
\right )}{\Gamma_\omega}\left |\frac{\Pi_{\rho^0\omega}}{m^2_\omega -
m^2_{\rho^0} - i\cdot m_\omega\left (\Gamma_\omega (m_\omega ) -
\Gamma_\rho^0(m_\omega )\right )}\right |^2\,.
\end{eqnarray}

As known \cite{goldhaber-69,gourdin-69,renard-70,achasov-78,achasov-92} one
can ignore $Im\left (\Pi_{\rho^0\omega}\right )$. Besides, the interference
pattern of the $\rho^0$ and $\omega$ mesons in the $e^+e^-\to\pi^+\pi^-$
reaction and in the $\pi^+\pi^-$ photoproduction on nuclei shows
\cite{goldhaber-69,gourdin-69,renard-70,achasov-78,achasov-92}
that $-Re\left (\Pi_{\rho^0\omega}\right )< 0$. So, taking into account
$B(\omega\to\pi^+\pi^-)=0.0221 \pm 0.003$ \cite{pdg-98} one gets
\begin{eqnarray}
\label{rerho0omega}
&& - Re\left (\Pi_{\rho^0\omega}\right )= -( 3.91\pm 0.27 )\cdot 10^{-3}
\mbox{GeV}^2\,.
\end{eqnarray}

From Eqs. (\ref{uspinV}), (\ref{kstar0kstarpm}) and (\ref{rerho0omega})
follows
\begin{eqnarray}
\label{rho0rhopmth}
&& m_{\rho^0} - m_{\rho^\pm} = 5.26 \pm 1.41 \ \ \mbox{MeV}\,.
\end{eqnarray}

This result is a puzzle. First, this mass splitting is considerable and
contrary to Eq. (\ref{rho0rhopm}).
Second, it is largely of electromagnetic origin also as the $\pi^\pm-\pi^0$
splitting but has the opposite sign. The $\rho^0$ meson is heavier than the
$\rho^\pm $ one!

If to consider the Eq. (\ref{uspinV}) as the linear one \cite{linear} then
$m_{\rho^0}-m_{\rho^\pm}=4.1\pm 1.2$ MeV and the situation does not change
essentially.

Certainly, it may be that corrections to Eq. (\ref{uspinV}) are important,
but the current theoretical understanding of the vector meson mass splitting
in the isotopical multiplets is far from being perfect, see, for example,
\cite{gasser-82,schechter-93,bijnens-96,gao-97}.

As for Eq. (\ref{rho0rhopm}), it stems from \cite{aleph-97} where the
$\tau^-\to\nu_\tau\pi^-\pi^0$ data \cite{aleph-97} are fitted in combination
with the $e^+e^-\to\pi^+\pi^-$ ones \cite{barkov-85}, which have the same,
excluding $\rho^0-\omega$ mixing, production mechanism. But a combined fit of
different experiments is open to a loss of sizable systematic errors.

That is why the problem of an alternative experimental measurement of the
$\rho^{\pm}-\rho^0$ mass splitting is ambitious enough. But this task is a
considerable challenge for it is practically meaningless to compare different
experiments with the different $\rho$ production mechanisms for the large
width of the $\rho$ meson .

The point is that our current knowledge of hadron production mechanisms is
far from being perfect and generally in the resonance region we have a
spectrum
\begin{eqnarray}
\label{spectrum}
&&\frac{dN}{dE}\sim \frac{f(E)}{\left ( E - E_R\right )^2 +
\frac{\Gamma^2}{4}}\,,
\end{eqnarray}
where $f(E)$ is a poorly varying in resonance region unknown function
\cite{f(e)} which can shift the visible peak up to a few MeV from $E_R$.

Really, let take into account two first terms of expansion of $f(E)$ in the
resonance region
\begin{eqnarray}
\label{fE}
&& f(E) = f_0 + \left ( E - E_R \right )f_1 + ...\,,
\end{eqnarray}
and let there be $(f_0/f_1)^2\gg (\Gamma/2)^2$, then the shift of the visible
peak
\begin{eqnarray}
\label{shift}
&& \Delta E_R = \frac{\Gamma^2}{8}\cdot\frac{f_1}{f_0}\,.
\end{eqnarray}
So, if $f_1=\pm f_0/(4.72\Gamma )= \pm 1.4f_0$ GeV$^{-1}$, $\Gamma = 151$ MeV,
then
\begin{eqnarray}
\label{shift}
&& \Delta E_R = \pm 4\,\mbox{MeV}\,.
\end{eqnarray}

Certainly, one can use other than $e^+e^-\to\pi^+\pi^-$ and
$\tau^-\to\nu_\tau\pi^-\pi^0$ different processes with the same $\rho^\pm$
and $\rho^0$ production mechanism, for example,
$a_1^-(1260)\to\rho^-\pi^0\to\pi^-\pi^0\pi^0$ and
$a_1^-(1260)\to\rho^0\pi^-\to\pi^+\pi^-\pi^-$ \cite{zaitsev-97}, the
advantage of which is the absence of the $\rho^0-\omega$ mixing. But in this
case the problem of different experimental systematic errors also exists.

It seems to us that the most adequate process for the aim under discussion
is the $\phi\to\rho^+\pi^- +\rho^-\pi^+ +\rho^0\pi^0\to\pi^+\pi^-\pi^0$
decay. Indeed, the charged and neutral $\rho$ mesons are produced in the one
reaction with the same mechanisms. Already now Spherical Neutral Detector
(SND) and Cryogenic Magnetic Detector-2 (CMD-2) at the $e^+e^-$ collider
VEPP-2M in Novosibirsk have collected $\sim 10^7\,\phi$ mesons each that is
$\sim 10^6\,\phi\to\rho\pi\to 3\pi$ decays each. With the $\phi$ factory
DA$\Phi$NE in Frascati, two orders of magnitude larger statistics will be
collected.

The differential cross section of the $e^+e^-\to\pi^+(k_+)\pi^-(k_-)\pi^0(k)$
reaction can be written in the symmetrical form \cite{gellmann-62,achasov-74}
\begin{eqnarray}
\label{eeto3pi}
&& \frac{d\sigma}{dm_+^2dm_-^2dm^2d\cos\vartheta_Nd\varphi}=\nonumber\\[1pc]
&& =\frac{\alpha^2|\vec k_+|^2|\vec k_-|^2\sin^2\vartheta_{+-}\sin^2
\vartheta_N}{128\pi^2s^2}|F|^2\delta (m_+^2+m_-^2+m^2-s-2m_{\pi^+}^2-
m_{\pi^0}^2 )\,,
\end{eqnarray}
where $m_+^2=(k_+ + k)^2\,, \,m_-^2=(k_- + k)^2\,,\, m^2=(k_+ + k_-)^2\,,\,
s=(k_+ + k_- + k)^2\,,\,\vartheta_N$ is the angle between the normal to the
production plane and the $e^+e^-$ beam direction in the center mass system,
$\vartheta_{+-}$ is the angle between the directions of the $\pi^+$ and
$\pi^-$  momenta in the center mass system.

The formfactor $F$ of the $\gamma^\ast\to\rho\pi$ decay with taking into
account the $\rho^0-\omega$ mixing has the form
\begin{eqnarray}
\label{formfactor}
&& F=A_\rho(s\,,\,m_+)\frac{2g_{\rho\pi\pi}(m_+)}{D_{\rho^+}(m_+)}
\exp\{i\cdot\delta(s\,,\,m_+)\} + A_\rho(s\,,\,m_-)\frac{2g_{\rho\pi\pi}(m_-)}
{D_{\rho^-}(m_-)}\exp\{i\cdot\delta (s\,,\,m_-)\} + \nonumber\\[1pc]
&& + A_\rho(s\,,\,m)\frac{2g_{\rho\pi\pi}(m)}{D_{\rho^0}(m)}\cdot\exp\{i\cdot
\delta (s\,,\,m)\}\left (1+\frac{A_\omega(s)}{A_\rho(s\,,\,m)}\cdot
\frac{\Pi_{\rho^0\omega}}{D_\omega (m)}\exp\{-i\cdot\delta (s\,,\,m)\}
\right )\,,
\end{eqnarray}
where $D_V(x)$ is a propagator of a $V$ meson, in the simplest case
$D_V(x)= m_V^2-x^2-i\cdot x\Gamma_V(x)$, $\Gamma_\rho (x)=
\left ( g^2_{\rho\pi\pi}(x)/6\pi\right )\left (q^3_\pi (x)/x^2\right )$, to a
good accuracy one can consider that propagators of the $\rho^\pm$ and
$\rho^0$ mesons differ by values of the masses $m_{\rho^\pm}^2$ and
$m_{\rho^0}^2$ only, $\delta (s\,,\,x)$ is a phase due to the triangle 
singularity ( the Landau anomalous thresholds ) \cite{achasov-94}.

At the $\phi$ meson energy $\left |A_\omega(s)/A_\rho (s\,,\,m)\right |
\simeq 0.02$, that is the $\rho^0-\omega$ mixing effects are negligible.
As the energy ( $\sqrt s$ ) increases the interference between terms in Eq.
(\ref{formfactor}) decreases and is inessential at $\sqrt s = 1.5-2$ GeV,
that is a circumstance favorable for the aim under consideration, but
the statistics in this energy region is poor, besides, the $\rho^0-\omega$
effects in this energy region are expected to be considerable
\cite{achasov-74,achasov-94}.

By itself the $J/\psi\to\rho\pi\to 3\pi$ decay stands. Generally speaking,
it is possible to select the adequate statistics in the future for
$B(J/\psi\to\rho\pi)=(1.28\pm 0.1)\cdot 10^{-2}$. The interference between
the terms in Eq. (\ref{formfactor}) is practically absent here, but the
$\rho^0-\omega$ mixing effects can essentially prevent the measurement of
the $\rho^\pm-\rho^0$ mass splitting ( $B(J/\psi\to\rho^0\pi^0=
(4.2\pm 0.5)\cdot 10^{-3}$ and $B(J/\psi\to\omega\pi^0=
(4.2\pm 0.6)\cdot 10^{-4}$ ), especially for the relative phase of the
amplitudes of the $J/\psi\to\rho^0\pi^0$ and $J/\psi\to\omega\pi^0$ decays is
unknown. The taking into account of the effects of the heavy $\rho^\prime$
mesons in the $J/\psi\to 3\pi$ decay one can find in \cite{achasov-97}.

We thank A.A. Kozhevnikov, G.N. Shestakov and A.M. Zaitsev for useful
discussions.

The present work was supported in part by the grant INTAS-94-3986.


\begin{references}
\bibitem{f(e)}
Efficiencies of registration can play a role of such functions for different
processes with the same $\rho$ production mechanism.
\end{references}
\end{document}